\documentclass[%
 reprint,
superscriptaddress,
 amsmath,amssymb,
 aps,
prb,citeautoscript
]{revtex4-2}

\usepackage{graphicx}
\usepackage{dcolumn}
\usepackage{bm}
\usepackage{lipsum}
\usepackage{siunitx}
\usepackage{placeins}
\usepackage{amstext}
\usepackage{mathtools} 
\usepackage{amsmath,amssymb}    
\usepackage{booktabs}
\usepackage{array}
\usepackage{xcolor}
\usepackage{tabularx}
\usepackage{ulem}
\usepackage{hyperref}
\usepackage{nameref}
\usepackage{float}
\usepackage{makecell}
\usepackage{caption}
\usepackage{booktabs}
\DeclareSIUnit{\rad}{rad}
\DeclareSIUnit{\deg}{deg}

\hypersetup{
        colorlinks   = true,
        citecolor    = blue,
        linkcolor    = blue}

\DeclareSIUnit\bar{bar}
\DeclareSIUnit\torr{Torr}

\makeatletter
\def\@email#1#2{%
 \endgroup
 \patchcmd{\titleblock@produce}
  {\frontmatter@RRAPformat}
  {\frontmatter@RRAPformat{\produce@RRAP{*#1\href{mailto:#2}{#2}}}\frontmatter@RRAPformat}
  {}{}
}%

\let\svthefootnote\thefootnote
\newcommand\freefootnote[1]{%
  \let\thefootnote\relax%
  \footnotetext{#1}%
  \let\thefootnote\svthefootnote%
}

\begin{document}
\title{Simplified Silicon Nitride Nanomembrane Circuits for van der Waals Integration}

\author{Tommaso Confalone}
\affiliation{Leibniz Institute for Solid State and Materials Research Dresden (IFW Dresden), 01069 Dresden, Germany}
\affiliation{Institute of Applied Physics, Dresden University of Technology, 01062 Dresden, Germany}

\author{Vasilisa Gerega}
\affiliation{Leibniz Institute for Solid State and Materials Research Dresden (IFW Dresden), 01069 Dresden, Germany}
\affiliation{Institute of Applied Physics, Dresden University of Technology, 01062 Dresden, Germany}

\author{Flavia Lo Sardo}
\affiliation{Leibniz Institute for Solid State and Materials Research Dresden (IFW Dresden), 01069 Dresden, Germany}
\affiliation{Institute of Materials Science, Dresden University of Technology, 01062 Dresden, Germany}

\author{Shreya Kumbhakar}
\affiliation{Leibniz Institute for Solid State and Materials Research Dresden (IFW Dresden), 01069 Dresden, Germany}
\affiliation{Institute of Solid State and Materials Physics, Dresden University of Technology, 01062 Dresden, Germany}

\author{Davide Massarotti}
\affiliation{Department of Electrical Engineering and Information Technology, University of Naples Federico II, I-80125 Naples, Italy}

\author{Francesco Tafuri}
\affiliation{Department of Physics, University of Naples Federico II, 80125 Naples, Italy}

\author{Shigeyuki Ishida}
\affiliation{Core Electronics Technology Research Institute, National Institute of Advanced Industrial Science and Technology (AIST), Tsukuba, Ibaraki 305-8565, Japan}

\author{Hiroshi Eisaki}
\affiliation{Core Electronics Technology Research Institute, National Institute of Advanced Industrial Science and Technology (AIST), Tsukuba, Ibaraki 305-8565, Japan}

\author{Kornelius Nielsch}
\affiliation{Leibniz Institute for Solid State and Materials Research Dresden (IFW Dresden), 01069 Dresden, Germany}
\affiliation{Institute of Applied Physics, Dresden University of Technology, 01062 Dresden, Germany}
\affiliation{Institute of Materials Science, Dresden University of Technology, 01062 Dresden, Germany}

\author{Golam Haider}
\thanks{g.haider@ifw-dresden.de}
\affiliation{Leibniz Institute for Solid State and Materials Research Dresden (IFW Dresden), 01069 Dresden, Germany}

\author{Nicola Poccia}
\thanks{nicola.poccia@unina.it}
\affiliation{Leibniz Institute for Solid State and Materials Research Dresden (IFW Dresden), 01069 Dresden, Germany}
\affiliation{Department of Physics, University of Naples Federico II, 80125 Naples, Italy}


\begin{abstract}
Two-dimensional (2D) materials and van der Waals (vdW) heterostructures provide an exceptional platform for engineering quantum devices, yet realizing their potential requires electrical integration without compromising the pristine properties of atomically thin crystals through conventional nanofabrication. Transferable circuitry addresses this challenge by decoupling circuit fabrication from device assembly, enabling electrical contacting without directly processing the active material. Here, we introduce SiN$_x$ nanomembrane (NMB) circuits realized through a simplified top-down strategy that reduces fabrication complexity, processing steps and specialized tools required by our previous bottom-up approach. As a stringent benchmark of material preservation, we electrically integrate a four-unit-cell-thick, optimally doped Bi$_2$Sr$_{2-x}$La$_x$CuO$_{6+\delta}$ (Bi2201) flake and observe a superconducting transition at T$_c^{inf}$\,$\sim$\,32\,K, close to the T$_c^{onset}$\,$\sim$\,34\,K measured by susceptibility in the parent crystals. The preservation of superconductivity demonstrates electrical integration of fragile layered materials without direct exposure to conventional cleanroom procedures, providing a versatile platform for integrating increasingly complicated vdW heterostructures, moiré materials, and hybrid quantum architectures.\\

\end{abstract}

\maketitle

\begin{figure*}
  \includegraphics[width=\textwidth]{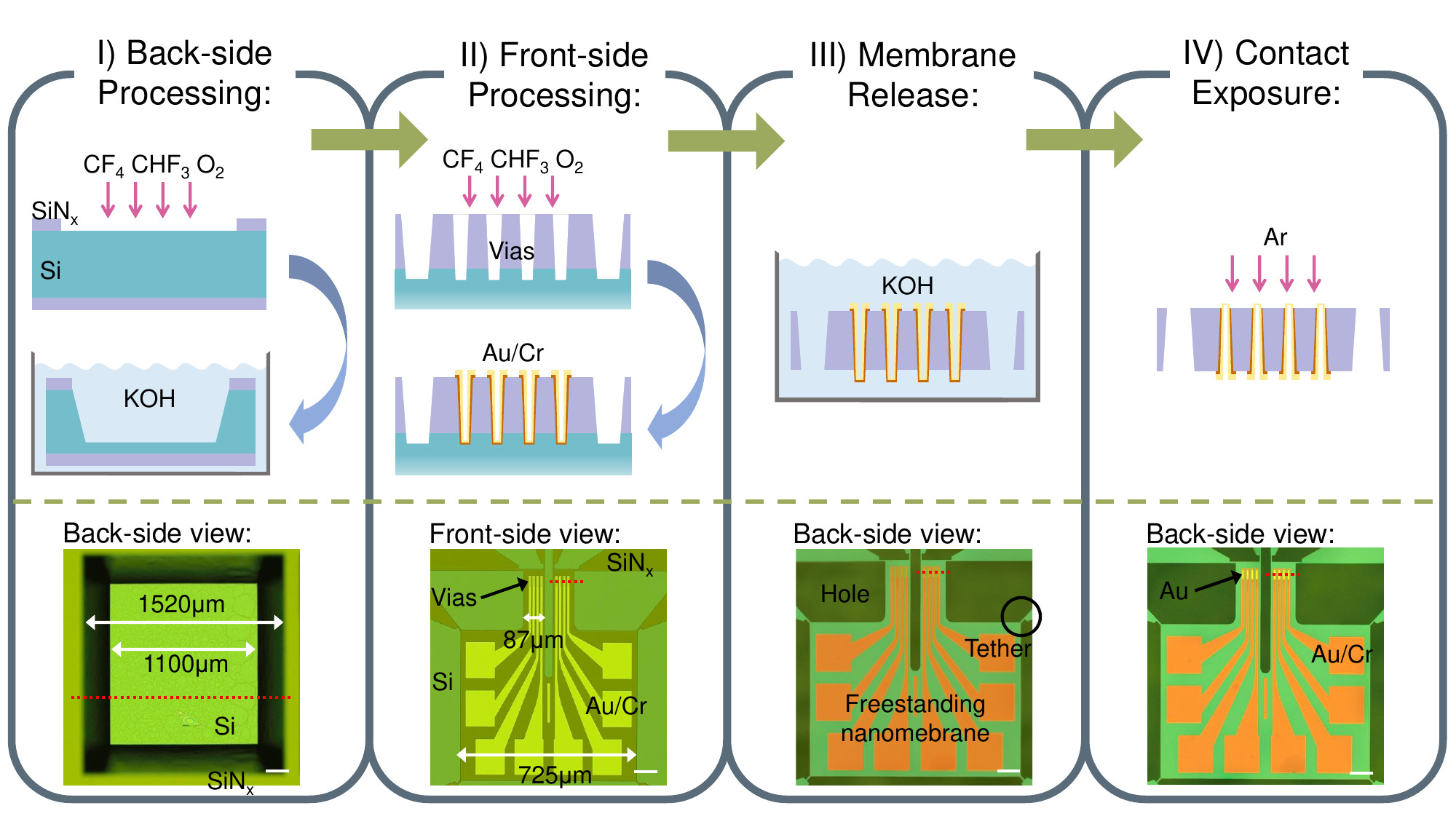}
  \caption{SiN$_x$ nanomembrane circuits fabrication process. Top: Schematic cross-sectional illustrations of the fabrication process, showing (I) backside array definition, (II) front-side SiN$_x$ etching and metallization, (III) membrane release by KOH etching, and (IV) contact exposure by Ar etching. Bottom: Corresponding optical images for each fabrication step. White scale bars correspond to $100\,\mu$m, and red dotted lines indicate the cross-sectional locations shown in the schematics above.}
  \label{fig1:fab}
\end{figure*}

\section{Introduction}
Two-dimensional (2D) materials provide a versatile platform for exploring novel quantum phenomena and developing new device concepts \cite{Novoselov2016, Geim2013, Liu2016}. In particular, the twist angle between adjacent layers provides a powerful tuning parameter \cite{Carr2017, Li2024, Andrei2021} for accessing emergent electronic phenomena, from moiré-induced correlated phases in twisted bilayer graphene \cite{Bistritzer2011, Cao2018, Cao2018bis} and transition-metal dichalcogenide heterostructures \cite{Regan2020, Wang2020, Xia2024} to probing a novel emerging superconducting order at the interface of twisted Bi$_2$Sr$_2$CaCu$_2$O$_{8+\delta}$ (Bi2212) heterostructures \cite{Zhao2023, Confalone2025, Basu2026, Martini2023, Lee2023}. These systems offer opportunities for both fundamental studies and applications in electronics, optoelectronics, sensing, and quantum technologies \cite{Wang2012, Confalone2025bis, Liu2019}.\\
The same structural characteristics that make 2D materials scientifically attractive also make their integration into functional devices particularly challenging. Because their active region can be only a few atomic layers thick, these materials are highly vulnerable to environmental exposure and fabrication-induced degradation. Conventional lithographic and metallization processes can introduce contamination, residues, structural disorder, and defects at the material surface and metal–2D interface \cite{Wang2022, Ma2024, Liu2018}. At the same time, low-resistance electrical contacts are essential for probing the underlying phenomena and require efficient carrier injection across interfaces whose properties are highly sensitive to bonding, contamination, and electrostatic environment \cite{Allain2015}. This challenge becomes particularly severe for sensitive materials approaching the monolayer or few-unit-cell limit \cite{Yu2019, Weinhold2021, Island2015}.\\
Several strategies have been developed to form high-quality electrical contacts to 2D materials. Geometrical approaches, such as one-dimensional edge contacts, enhance electrode–crystal coupling while overcoming limitations of conventional basal-plane contacts \cite{Wang2013, Jain2019, Jena2014}. Electrostatic engineering, including local gating and contact-induced doping, modifies the carrier density near the interface and reduces injection barriers \cite{Liao2019, Duflou2023, Das2013}. More recently, vdW contacts have enabled electrically active interfaces while minimizing strong chemical bonding and associated damage to the underlying crystal \cite{Wang2022,Liu2018}. However, the implementation of these approaches can require processing conditions, involving thermal, physical and chemical treatments, incompatible with the preservation of highly sensitive materials. The challenge is therefore not simply to optimize contact, but to preserve the material throughout the entire process.\\
To address this fundamental limitation, we previously developed a methodology that decouples electrical-circuit fabrication from processing of the active 2D material \cite{Saggau2023, Shokri2025, Martini2023bis}. In this approach, the desired circuit is fabricated on a SiN$_x$ nanomembrane (NMB), which acts as a transferable body for subsequent integration with the device under investigation. This strategy transforms circuit fabrication from an on-device process into a separate fabrication and integration step, thereby avoiding the direct exposure of fragile 2D materialS to conventional processing. Using this methodology, we demonstrated electrical contacts to ultrathin Bi2212 flakes \cite{Saggau2023, Shokri2025} and twisted Bi2212 heterostructures \cite{Confalone2025, Confalone2025tris}. \\
Despite these successes, the previously developed methodology relies on a complex fabrication sequence and specialized processing tools, limiting its accessibility and broader adoption \cite{Saggau2023}. Here, we present a streamlined implementation of the same transferable-circuit concept that maintains the core idea of separating circuit fabrication from active-material processing, while reducing process complexity and tool requirements. We demonstrate the approach by integrating electrical contacts with a four-unit-cell-thick, optimally doped Bi$_2$Sr$_{2-x}$La$_x$CuO$_{6+\delta}$ (Bi2201) flake and measuring its temperature-dependent resistance, showing the preservation of superconductivity in this ultrathin limit. With increasing efforts to fabricate complex vdW heterostrcutures \cite{Park2022, Khalaf2019}, the decoupling of circuit fabrication from active material processing provides a novel strategy to integrate the sensitive 2D interfaces into functional devices.\\

\begin{figure*}
  \includegraphics[width=\textwidth]{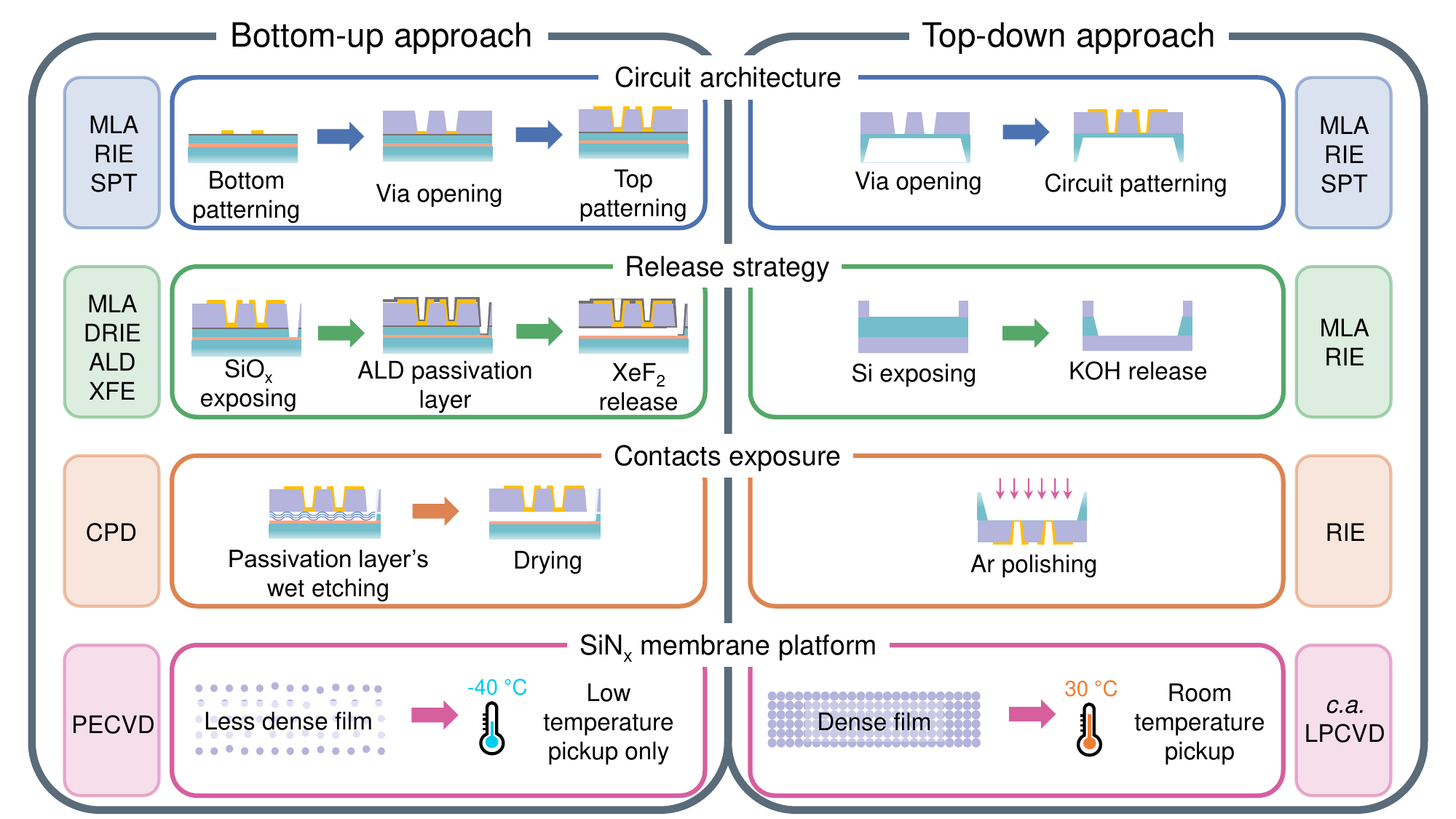}
  \caption{Schematic comparison of the fabrication workflows for the bottom-up and top-down approaches. The process flows are divided into four conceptual categories, with the required fabrication tools listed for each category on both sides. MLA = maskless aligner; (D)RIE = (deep) reactive ion etching; SPT = sputtering tool; ALD = atomic layer deposition; XFE = XeF$_2$ etcher; CPD = critical point dryer; PECVD = plasma-enhanced chemical vapor deposition; c.a. LPCVD = commercially available low-pressure chemical vapor deposition.}
  \label{fig2:comparison}
\end{figure*}

\section{Methods and Results}

\subsection{Fabrication workflow of the top-down membrane platform}
The fabrication strategy for the SiN$_x$ NMB circuits developed in this work is based on a top-down procedure in which the circuit is directly patterned onto a commercially available 400\,nm\,SiN$_x$\,/\,300\,$\mu$m\,Si\,/\,400\,nm\,SiN$_x$ wafer. Unlike the previously developed bottom-up process \cite{Saggau2023}, the structural SiN$_x$ layer of the transferable membrane is already present at the beginning of fabrication, thereby reducing the number of processing steps and eliminating in-house SiN$_x$ deposition and intermediate handling. The overall process, outlined in Figure\,\ref{fig1:fab}, consists of four stages: I) backside processing, II) frontside processing, III) membrane release, and IV) contact exposure.\\
The use of a pre-existing thin film as a transferable circuit platform introduces two main challenges: (i) releasing the membrane from the substrate while maintaining its structural integrity and (ii) establishing robust vertical electrical connections across the membrane. The first is addressed through sacrificial-layer removal, a common strategy for fabricating freestanding thin films \cite{Lu2016, Chiabrera2022}, while the second requires through-membrane vias and suitable metallization.\\
Backside processing defines access to the sacrificial Si layer while preserving sufficient mechanical support for subsequent processing. An array of square openings ($\sim$\,1500\,x\,1500\,$\mu$m$^2$) is patterned in the backside SiN$_x$ layer via photolithography and reactive ion etching (RIE), exposing the underlying Si. Subsequent anisotropic etching in 40\% KOH solution at 80\,$^\circ$C forms V-shaped cavities with sidewalls at 54.7\,$^\circ$ \cite{Madou2018}, resulting in frontside membrane windows of $\sim$\,1100\,x\,1100\,$\mu$m$^2$. A residual Si thickness of 4–10\,$\mu$m is intentionally preserved to allow the subsequent metal deposition while the membrane remains mechanically supported. Moreover, the partially thinned substrate permits alignment of the frontside pattern with the backside openings in the following steps.\\
Frontside processing defines both the membrane geometry and the through-membrane holes (vias) using aligned photolithography and RIE. Rectangular vias with widths of 2–4\,$\mu$m and lengths of 4–30\,$\mu$m are chosen to balance mechanical stability and electrical performance. Subsequent sputter deposition defines the circuit layout, simultaneously forming the in-plane circuitry and establishing vertical electrical interconnections through metal coating of the via bottoms and sidewalls.\\
The metallization stack used for the electrical leads consists of 1\,nm\,Cr/80\,nm\,Au/15\,nm\,Cr/80\,nm\,Au. This multilayer structure is designed to provide adhesion, mechanical stability, and chemical robustness throughout the fabrication process. The initial Cr layer promotes adhesion to the SiN$_x$ surface while remaining sufficiently thin to allow its selective removal in the final step. The intermediate Cr layer enhances mechanical rigidity within the vias \cite{Mulloni2009, Nemirovsky1978}, preventing deformation of the metal, while the Au layers provide low-resistance electrical pathways and protect the underlying layers during KOH etching.\\
The membrane is released by a second KOH etching step that fully removes the remaining Si. Following this etching step, the membrane remains connected to the surrounding SiN$_x$ frame only through narrow tethers, typically located at the corners of the structure. These tethers, with dimensions of approximately 5\,$\mu$m in width and 20\,$\mu$m in length, provide temporary mechanical support while preserving transferability. Their position and geometry are designed to control the fracture location, enabling predictable and reproducible detachment of the membrane during transfer.\\
A final Ar plasma etch removes the exposed 1-nm-thick Cr layer at the bottom of the vias. The etching parameters are optimized to ensure complete Cr removal while minimizing any increase in the surface roughness of the exposed Au.\\
These steps enable controlled membrane release, robust vertical electrical interconnections, and mechanical stability throughout fabrication while simplifying the overall process. Further details on the individual fabrication steps are provided in the Supplementary Information.\\

\subsection{Comparison with the bottom-up fabrication process}
The previously developed methodology relies on a bottom-up fabrication scheme in which the transferable SiN$_x$ NMB and electrical circuitry are assembled through sequential deposition and patterning steps. Although this strategy enables complex circuit architectures, its fabrication sequence involves several specialized processes. To provide a direct comparison with the streamlined top-down process developed in this work, we briefly summarize the main steps of the previous fabrication procedure.\\
Bottom contacts are first defined by lithography and metallization on an AlO$_x$ layer deposited by atomic layer deposition (ALD) on a silicon-on-insulator (SOI) substrate. A 400-nm-thick SiN$_x$ layer is then deposited by chemical vapor deposition (CVD), followed by reactive ion etching (RIE) to open vias at selected locations and at the same time defining the NMB geometry. Top contacts are subsequently defined by lithography and metallization to connect the bottom contacts through the SiN$_x$ layer. Finally, a protective AlO$_x$ layer is deposited, and XeF$_2$ etching and critical-point drying (CPD) are used to release and dry the SiN$_x$ NMB. Further details of the process are reported in Ref.\,\cite{Saggau2023}.\\
To enable a systematic comparison, the two fabrication workflows are considered across four key aspects: circuit architecture, release strategy, contact exposure, and SiN$_x$ NMB platform. The comparison is schematically summarized in Figure\,\ref{fig2:comparison}.\\
The main difference in circuit architecture lies in the number and sequence of fabrication steps required to realize the electrical circuitry. In the bottom-up process, circuit definition proceeds through multiple stages, including delineation of bottom contacts, deposition of the SiN$_x$ layer, via opening, and subsequent fabrication of the top circuitry. In contrast, the top-down process leverages the pre-existing SiN$_x$ layer to define the circuitry in a single aligned metallization step following via opening. In this configuration, the vias simultaneously provide access to the bottom surface and establish vertical electrical interconnections, allowing the transferable circuit to be realized in a more compact process flow. The bottom-up process, however, provides greater design flexibility because the bottom and top circuitry can be patterned independently.\\
The two approaches differ substantially in how the SiN$_x$ membrane is released. In the bottom-up process, the 2-$\mu$m-thick Si layer of the SOI substrate acts as the sacrificial layer. Access to this layer requires deep reactive ion etching (DRIE) to reach the buried oxide, followed by XeF$_2$ etching for selective Si removal. Moreover, an AlO$_x$ layer deposited by ALD is required to protect the membrane during this process. In contrast, the top-down process uses a double-side-coated Si wafer (SiN$_x$/Si/SiN$_x$), in which the 300-$\mu$m-thick Si substrate itself serves as the sacrificial layer and is removed by wet etching in KOH solution. Access to the Si is obtained by opening the backside SiN$_x$ layer using RIE. This sequence eliminates the need for DRIE, XeF$_2$ etching, and protective capping layers.\\
The release and contact-exposure procedures are coupled through the sequence required to access the bottom contacts. In the bottom-up process, the bottom Au layer is exposed by removing the protective AlO$_x$ capping layer through wet etching. Because the released membrane is separated from the substrate by only $\sim$\,2$\mu$m, capillary forces during drying can cause membrane collapse, requiring CPD. In the top-down process, the membrane is flipped after release, allowing the exposed 1-nm-thick Cr layer at the bottom of the vias to be selectively removed by Ar plasma etching. This sequence eliminates the need for CPD while providing direct access to the contact interface.\\
The two processes also employ SiN$_x$ films deposited by different techniques. The bottom-up process uses PECVD SiN$_x$, whereas the top-down process relies on commercially available LPCVD SiN$_x$/Si/SiN$_x$ wafers. Under otherwise comparable membrane thickness and geometry, we observe a marked difference in membrane handling: the LPCVD membranes can be picked up reliably at room temperature, whereas the membranes fabricated using the previous process required low-temperature transfer conditions.\\
A qualitative summary of the main differences between the two fabrication workflows is provided in Table\,\ref{tab1}. The top-down process requires fewer processing stages and specialized tools, while its main limitations are reduced circuit-design flexibility and increased roughness at the contact interface resulting from via formation and metallization.\\

\begin{table*}
\caption{Qualitative comparison between the bottom-up and top-down fabrication approaches.}
\vspace{3mm}
\begin{tabular}{l|ccc}
\centering
 & \textbf{Bottom-up approach} & & \textbf{Top-down approach} \\\\
\hline
\\
%\textbf{2D materials integration} & \makecell{*** \\ (In-process integration easier)} & & \makecell{** \\ (less prone to in-process integration)} \\
%\\
\textbf{Process complexity \& tools} & \makecell{* \\ (multi-step, tool-intensive)} & & \makecell{*** \\ (streamlined, minimal tools)} \\
\\
\textbf{Process reliability} &  \makecell{** \\ (multi-step failure risk)} & & \makecell{*** \\ (minimized failure risk)} \\
\\
\textbf{Handling \& transfer easiness} &  \makecell{* \\ (low-temperature pickup required)} & & \makecell{*** \\ (room temperature pickup possible)} \\
\\
\textbf{Circuit design flexibility} & \makecell{*** \\ (high geometric freedom)} & & \makecell{** \\ (layout constraints)} \\
\\
\textbf{Contact roughness} & \makecell{*** \\ (flat surface)} & & \makecell{** \\ (rougher surface)} \\
\\
\end{tabular}
\vspace{3mm}
\caption*{Legend: *** = excellent, ** = good, * = limited.}
\label{tab1}
\end{table*}

\subsection{Contact interface engineering}
The top-down process introduces additional sources of surface roughness at the contact interface compared with the intrinsically flat contacts of the bottom-up platform. A central objective was therefore to minimize the morphological changes introduced during fabrication and to obtain a contact surface suitable for integration with 2D materials and vdW heterostructures. The resulting contact interface is primarily determined by three fabrication steps: I) via opening by RIE, which defines the initial contact geometry; II) KOH etching for membrane release, which can affect the integrity of the metal contacts; and III) Ar plasma treatment, which exposes the contact surface by removing the underlying Cr layer. AFM and SEM measurements were used to characterize the effects of these steps, as shown in Figure\,\ref{fig3:AFM_SEM}.\\
During via opening by RIE, two aspects are critical: the bottom profile of the etched vias and the roughness of the exposed Si surface beneath the SiN$_x$ layer. These features determine the initial conditions for subsequent metal deposition and ultimately influence the morphology of the contact interface. See Supplementary Information Figure\,S3 for SEM images of the via at different stages of the fabrication. Ideally, the via bottom should be flat and smooth to provide a uniform contact surface for the subsequent integration of 2D materials and vdW heterostructures. Because SiN$_x$ is insulating, charge accumulation during plasma etching can deflect ions near patterned edges, producing microtrenching and an inverted U-shaped via bottom (Figure\,\ref{fig3:AFM_SEM}(a)) \cite{Donnelly2013}. Increasing the process pressure reduces ion anisotropy and mitigates this effect \cite{Seta2002}, resulting in a flatter via bottom. The surface roughness is instead controlled by adjusting the RIE gas composition, yielding an RMS roughness of $\sim$\,1.6\,nm over a 5x5\,$\mu$m$^2$ area of the exposed Si surface (Figure\,\ref{fig3:AFM_SEM}(b)). CF$_4$ is used as the primary SiN$_x$ etchant, while CHF$_3$ and O$_2$ are added to control the etch rate and improve surface smoothness.\\
During the final KOH release step, the etching time must be carefully controlled to ensure complete removal of the remaining Si while preserving the integrity of the metallic contacts. Excessive exposure can compromise the Cr adhesion layer, leading to partial delamination of the Au and resulting in a non-uniform contact surface. An AFM scan of a contact subjected to prolonged KOH etching is shown in Figure\,\ref{fig3:AFM_SEM}(c), where the delaminated regions appear as holes in the Au surface.\\
\subsection{Integration with an ultrathin van der Waals superconductor}
\begin{figure*}
  \includegraphics[width=\textwidth]{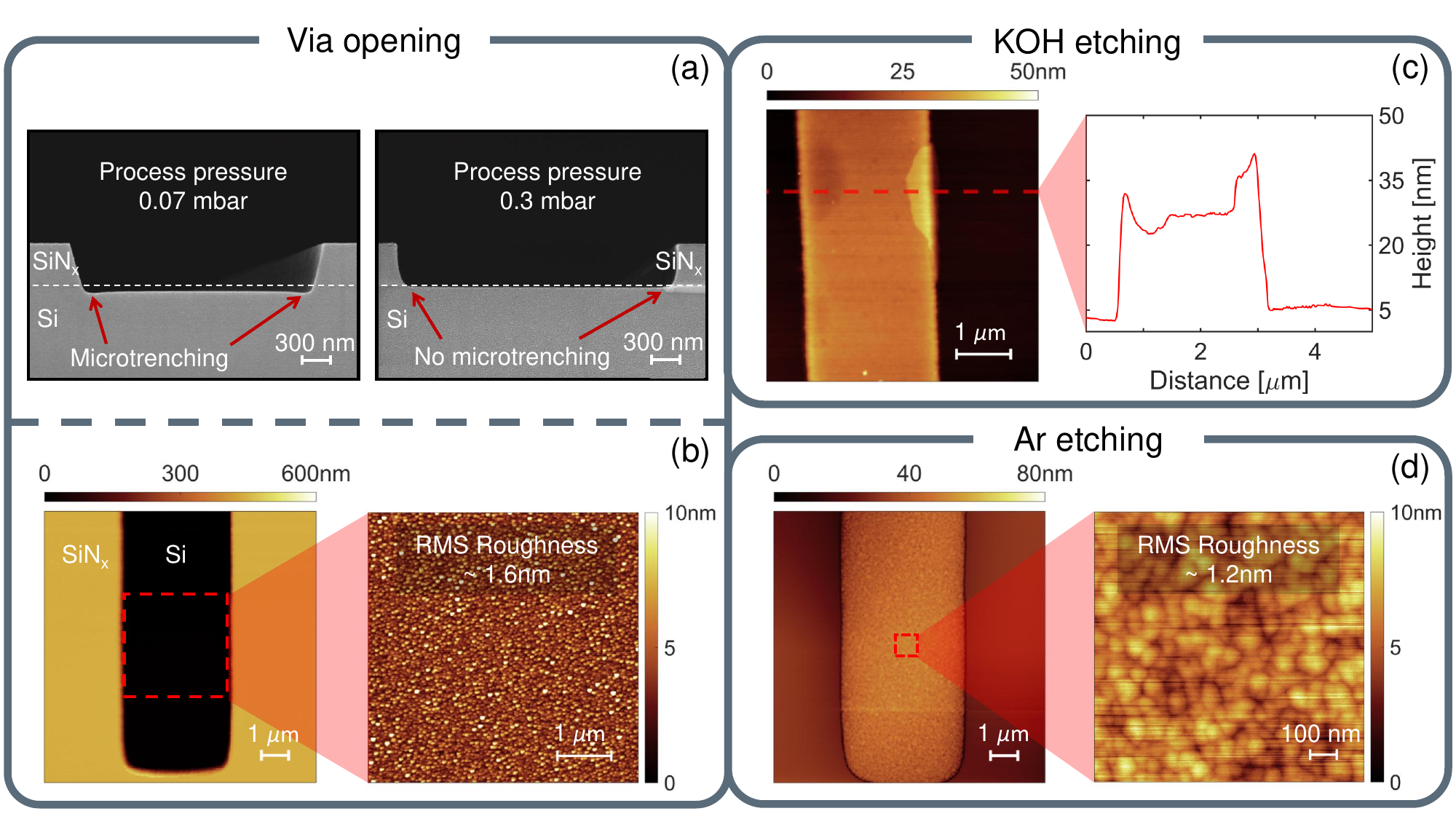}
  \caption{SEM and AFM characterization of the contact engineering process. (a) SEM images comparing SiN$_x$ etching with and without microtrenching. (b) AFM image of the via after optimized SiN$_x$ etching. (c) AFM image of a contact after over-etching in KOH, showing holes and surface non-uniformity caused by partial detachment of the Au/Cr layer. (d) AFM image of a contact after Ar etching to remove the Cr layer. The etching conditions were optimized to maintain a surface roughness comparable to that before Cr removal.}
  \label{fig3:AFM_SEM}
\end{figure*}
The final Ar plasma treatment exposes the contacts by selectively removing the 1-nm-thick Cr layer at the bottom of the vias. This step requires a balance between complete Cr removal and preservation of the underlying Au surface, since prolonged or overly aggressive plasma exposure can increase surface roughness. To address this trade-off, the etching is divided into a high-power step for efficient Cr removal followed by a low-power treatment to minimize damage to the exposed Au. The resulting contact surface is characterized by AFM in Figure\,\ref{fig3:AFM_SEM}(d), showing that the optimized etching conditions preserve the morphology of the contact surface established after via opening. See Supplementary Information Figure\,S4 instead for images of Cr removal via wet etching.\\
To validate the top-down fabrication strategy, we investigate its ability to electrically integrate an ultrathin device while preserving the intrinsic properties of the active material. As a benchmark system, we selected Bi2201, a vdW superconductor whose superconducting properties are intimately linked to the concentration and spatial ordering of interstitial oxygen \cite{Fratini2010,Poccia2011}. The distribution of interstitial oxygen in real space directly influences the electronic structure and superconducting state, making the material particularly sensitive to perturbations that alter its oxygen content or ordering. This sensitivity becomes especially pronounced in the ultrathin limit, making Bi2201 a stringent benchmark for a contact-integration strategy designed to minimize environmental exposure and fabrication-induced modifications.\\
\begin{figure*}
  \includegraphics[width=\textwidth]{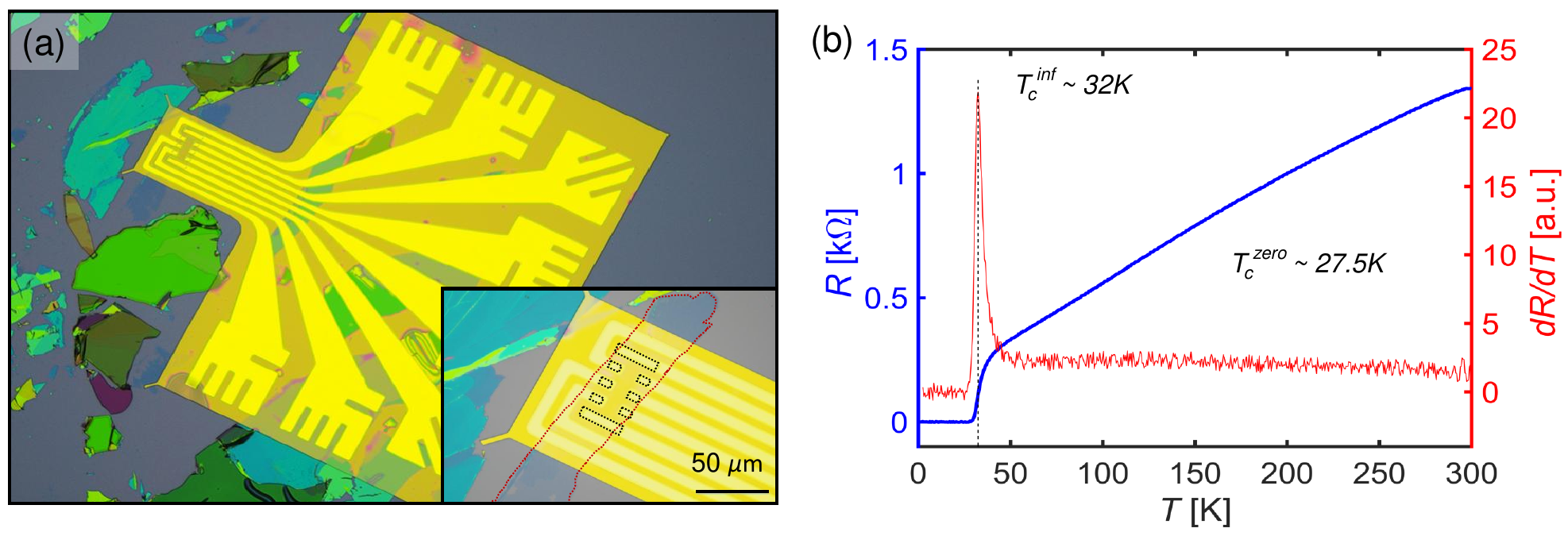}
  \caption{Electrical contact integration of a four-unit-cell-thick, optimally doped Bi2201 flake. (a) Optical image of the device. Inset: Magnified view of the flake (red dotted line) and contacts (black dotted lines). (b) Temperature dependence of the resistance, $R(T)$, and its derivative, $dR/dT$, from $300\,$K to $3\,$K. The superconducting transition temperatures defined from the inflection point of $R(T)$, $T_c^{inf}$, and from the onset of zero resistance, $T_c^{zero}$, are indicated.}
  \label{fig4:transport}
\end{figure*}
Bi2201 crystals were mechanically exfoliated onto Si/SiO$_2$ substrates using the Scotch-tape method, with the exfoliation stage maintained at -40\,$^\circ$C. The SiN$_x$ NMB carrying the electrical contacts was subsequently transferred onto the selected flake within 1\,min, while maintaining the same temperature. Both steps were performed in an inert-atmosphere glovebox with a H$_2$O concentration of approximately 10 ppb. The substrates were pretreated with oxygen plasma and baked overnight at 150\,$^\circ$C prior to exfoliation, to minimize environmental degradation and preserve the interstitial oxygen ordering in the ultrathin Bi2201. AFM measurements done after complete electronic characterization (see Supplementary Information Figure\,S1) confirm a thickness of approximately four-unit-cell ($\sim$9–10 nm), placing the device in the ultrathin regime where preservation of the material properties is particularly challenging \cite{Yu2019, Sandilands2010, Sardo2025}. The optimally doped Bi2201 single crystals were grown by the floating-zone method \cite{Eisaki2004, Ono2003}. \\ 
Following transfer of the NMB, the device was bonded outside the glovebox and loaded into the cryostat, resulting in an estimated air exposure of $\sim$20\,min. During this atmospheric exposure time, the SiN$_x$ NMB passivates the atomically thin Bi2201, minimizing its direct exposure to the ambient environment. Figure\,\ref{fig4:transport} shows the temperature-dependent resistance $R(T)$ of the resulting device, which exhibits a clear superconducting transition. The inflection point of the $R(T)$ curve yields T$_c^{inf}$\,$\sim$\,32\,K, close to the T$_c^{onset}$\,$\sim$\,34\,K determined from susceptibility measurements of the parent Bi2201 crystals (see Supplementary Information Figure S2). The preservation of a T$_c^{inf}$\,$\sim$\,32\,K comparable to that of the starting crystal prove that the membrane transfer and contact-integration process retains superconductivity of Bi2201 down to the ultrathin limit.\\
These results demonstrate that the top-down platform can preserve superconductivity in ultrathin Bi2201 while enabling electrical integration without exposing the active material to conventional fabrication processes.\\

\section{Discussion}
The results presented above show that the top-down SiN$_x$ NMB platform preserves the essential functionality of the previously developed transferable-circuit methodology while substantially simplifying its fabrication workflow. The reduction in process complexity is compatible with electrical integration of ultrathin, highly sensitive devices, while reducing reliance on specialized tools and processing steps.\\
This simplification also shifts the main fabrication challenge toward the contact interface. The top-down process requires the electrical contacts to be formed through etched vias, making their final morphology dependent on the combined effects of RIE, KOH release, and Ar plasma treatment. The optimization of these steps allows the resulting interface to remain suitable for electrical integration despite the additional processing required to expose the contacts.\\
The successful integration of four-unit-cell-thick Bi2201 flake further validates the platform, given the strong sensitivity of Bi2201 superconductivity to interstitial oxygen ordering. The observation of a transport transition close to that of the parent crystals indicates that the transfer and integration process preserves the superconducting properties of the material in the ultrathin limit. This provides a stringent demonstration of the compatibility of the platform with a particularly sensitive vdW superconductor.\\
An additional advantage of the top-down implementation arises from the timing of circuit definition within the fabrication sequence. In the bottom-up process, the circuit geometry is established from the beginning through the definition of the bottom contacts, requiring the device geometry to be tailored to the membrane. In the top-down process, the backside processing can instead be completed independently of the heterostructure. The circuit can then be tailored once the device geometry is known, enabling rapid integration without restarting the membrane fabrication. This capability is particularly valuable for fragile materials, for which minimizing the time between device assembly and electrical integration can reduce environmental exposure.\\
The top-down implementation introduces trade-offs that define its present applicability range. Compared with the bottom-up methodology, the circuit geometry is less independently tunable because the top and bottom electrical pathways are linked through the SiN$_x$ membrane etching, whereas they can be patterned independently in separate steps in the bottom-up process. The use of KOH for membrane release may further constrain compatibility with specific metallization materials. The contact interface is likewise system dependent: while the additional roughness introduced by the top-down process does not prevent electrical integration in the present devices, more demanding heterostructures or smaller device dimensions may require further interface engineering to achieve optimal electrical transparency \cite{Wong2024, Li2025}.\\
Beyond Bi2201, decoupling circuit fabrication from heterostructure processing could provide a useful strategy for increasingly complex quantum architectures. Fabricating and optimizing the circuitry independently, followed by integration after heterostructure assembly, may facilitate the implementation of fragile multilayer, twisted, and hybrid structures while reducing constraints on their processing. Transferable circuitry could therefore extend the applicability of this integration strategy beyond individual sensitive materials toward more complex assembled quantum heterostructures.\\

\section{Conclusion}
In conclusion, we developed a simplified top-down implementation of the transferable SiN$_x$ NMB methodology for integrating ultrathin vdW superconductors and, more broadly, fragile vdW systems. By retaining the key functionality of the previously developed bottom-up implementation while reducing fabrication complexity and specialized tool requirements, the new process provides a more accessible route to interface-preserving electrical integration. The preservation of superconductivity in a four-unit-cell-thick, optimally doped Bi2201 flake demonstrates the compatibility of this strategy with highly sensitive layered superconductors. As vdW heterostructures and their associated device architectures continue to increase in complexity, decoupling circuit fabrication from active-material processing establishes a versatile route toward integrating fragile quantum materials with increasingly sophisticated electrical circuitry.\\

\noindent\textbf{Acknowledgements.} N.P. acknowledges the partial funding by the European Union (ERC-CoG, 3DCuT, 101124606), the Deutsche Forschungsgemeinschaft (DFG, German Research Foundation): DFG 460444718, DFG 512734967, DFG 452128813, DFG 539383397, DFG 572638824. G.H. acknowledges financial support through start-up funding from Leibniz IFW Dresden. F.T. and D. M. would like to acknowledge the PNRR MUR Project PE0000023 NQST. The sample preparation was supported by JSPS KAKENHI (Grant No. JP19H05823). The authors are grateful to Heiko Reith and Nicolas Perez for providing access to cleanroom and cryogenic facilities, respectively.\\

\noindent\textbf{Author contributions.} N.P and G. H. supervised the project. T.C., G.H. and N.P. conceived and designed the experiment. T.C. performed the experiments and analyzed the data with the contribution of V.G. and F.L.S. S.I. and H.E. provided the cuprate crystals. T.C., V.G., F.L.S., S.K. and N.P. discussed the fabrication procedure. T.C., G.H. and N.P. discussed the results. T.C., G.H., K.N. and N.P. wrote the manuscript. All authors discussed the manuscript.\\

\noindent\textbf{Declaration of Interest} All authors declare no conflict of interest.\\

\noindent\textbf{Data Availability Statement} The data that support the findings of this study are available from the corresponding authors upon reasonable request.

\bibliographystyle{ieeetr}
\bibliography{bibliography}

\end{document}